\documentclass[article,preprint,groupedaddress,11pt]{revtex4}
\usepackage{epsfig}
\usepackage{graphicx}
\usepackage{amssymb}

\begin{document}
\title{Hawking radiation may violate the Penrose cosmic censorship conjecture}
\author{Shahar Hod}
\affiliation{The Ruppin Academic Center, Emeq Hefer 40250, Israel}
\affiliation{ } \affiliation{The Hadassah Institute, Jerusalem
91010, Israel}
\date{\today}
\centerline {\it This essay received an Honorable Mention from the
Gravity Research Foundation 2019}

\begin{abstract}
\ \ \ We analyze the Hawking evaporation process of
Reissner-Nordstr\"om black holes. It is shown that the
characteristic radiation quanta emitted by the charged black holes
may turn near-extremal black-hole spacetimes into horizonless naked
singularities. The present analysis therefore reveals the intriguing
possibility that the semi-classical Hawking evaporation process of
black holes may violate the fundamental Penrose cosmic censorship
conjecture.
\newline
\newline
Email: shaharhod@gmail.com
\end{abstract}
\bigskip
\maketitle

{\it Introduction}. ---\ \ The seminal work of Hawking
\cite{Haw1,Haw2} has revealed the intriguing fact that
semi-classical black-hole spacetimes are characterized by filtered
black-body emission spectra with well defined thermodynamic
properties \cite{Haw1,Haw2,Bekt}. Soon after his groundbreaking
discovery, Hawking noted that the thermally distributed black-hole
radiation spectrum may contradict the fundamental quantum principle
of a unitary time evolution \cite{Haw1,Haw2}. The incompatibility of
general relativity and quantum mechanics, as reflected by the
Hawking black-hole radiation phenomenon, is certainly one of the
most important open problems in modern physics.

In the present essay we would like to discuss another disturbing
feature of the Hawking evaporation mechanism of black holes. In
particular, we shall explicitly prove that the Hawking
semi-classical radiation process may turn a near-extremal
Reissner-Nordstr\"om (RN) black-hole spacetime into an horizonless
naked singularity which violates the black-hole condition $Q\leq M$
\cite{NoteQM,Noteunit}. Thus, our analysis, to be presented below,
suggests that the Hawking radiation of black holes may violate the
fundamental Penrose cosmic censorship conjecture \cite{Pen1,Pen2}
which asserts that spacetime singularities are always hidden behind
event horizons inside black holes.

{\it The Hawking evaporation process of near-extremal
Reissner-Nordstr\"om black holes}. ---\ \ We consider the
semi-classical Hawking evaporation process of RN black holes in the
near-extremal regime \cite{Notepqm}
\begin{equation}\label{Eq1}
0\leq\Delta\equiv M-Q\ll M\  .
\end{equation}
Note that, for a given value of the electric charge $Q$, a minimal
mass (extremal) black-hole spacetime is characterized by the simple
relation $\Delta=0$ [that is, $M_{\text{min}}(Q)=Q$].

Our analysis is based on the following two well-known facts
\cite{Pageev,Hodec}:
\newline
(1) For near-extremal RN black holes in the large-mass regime
\begin{equation}\label{Eq2}
M\gg{{e\hbar}\over{\pi m^2_e}}\  ,
\end{equation}
the quantum emission of massive charged fields (here $m_e$ and $e$
are respectively the proper mass and the electric charge of the
elementary positron field) is exponentially suppressed as compared
to the Hawking quantum emission of massless neutral fields.
\newline
(2) In addition, due to the partial back-scattering of the emitted
field quanta by the centrifugal barrier which surrounds the black
holes, the neutral sector of the Hawking radiation spectra of
spherically symmetric black holes is dominated by
electromagnetic field quanta with unit angular momentum
\cite{Notel1}.

As we shall explicitly show below, these two facts may allow a
near-extremal charged RN black hole in the regime (\ref{Eq2}) to
jump over extremality by emitting a characteristic neutral Hawking
quantum which reduces the mass of the black hole without reducing
its electric charge.

For one bosonic degree of freedom, the Hawking radiation power of
non-rotating black holes is given by the simple integral relation
\cite{Haw1}
\begin{equation}\label{Eq3}
P={{\hbar}\over{2\pi}}\sum_{l,m}\int_0^{\infty} {{\Gamma
\omega}\over{e^{\hbar\omega/T_{\text{BH}}}-1}}d\omega\  .
\end{equation}
Here
\begin{equation}\label{Eq4}
T_{\text{BH}}={{\hbar(M^2-Q^2)^{1/2}}\over{2\pi[M+(M^2-Q^2)^{1/2}]^2}}\
\end{equation}
is the Bekenstein-Hawking temperature of the RN black hole which, in
the near-extremal regime (\ref{Eq1}), is characterized by the strong
inequality ${{MT_{\text{BH}}}/{\hbar}}\ll1$. The integer parameters
$l$ and $m$ are respectively the spheroidal and axial angular
harmonic indices of the emitted field mode and the
frequency-dependent greybody factors $\Gamma=\Gamma_{lm}(\omega)$ in
(\ref{Eq3}) quantify the partial back-scattering of the field modes
by the curved spacetime outside the black-hole horizon \cite{Haw1}.

The familiar black-body (thermal) factor
$\omega/(e^{\hbar\omega/T_{\text{BH}}}-1)$ which appears in the
Hawking expression (\ref{Eq3}) for the black-hole bosonic radiation
power implies that the corresponding emission spectrum has a
characteristic peak at the dimensionless emission frequency
\begin{equation}\label{Eq5}
M\omega^{\text{peak}}\sim {{MT_{\text{BH}}}\over{\hbar}}\ll1\  ,
\end{equation}
in which case the frequency dependent greybody factors
$\Gamma_{lm}(\omega)$ are given by the simple low-frequency
analytical expression \cite{Pageev,Hodec}
\begin{equation}\label{Eq6}
\Gamma_{1m}={1\over9}\epsilon^8\nu^4(1+\nu^2)(1+4\nu^2)\cdot[1+O(M\omega)]\
,
\end{equation}
where
\begin{equation}\label{Eq7}
\epsilon\equiv {{2(M^2-Q^2)^{1/2}}\over{M+(M^2-Q^2)^{1/2}}}\ \ \ \
\text{and}\ \ \ \ \nu\equiv {{\hbar\omega}\over{4\pi
T_{\text{BH}}}}\  .
\end{equation}
Substituting (\ref{Eq6}) into the semi-classical Hawking relation
(\ref{Eq3}), one finds the compact expression \cite{Noteerx}
\begin{equation}\label{Eq8}
P={{\hbar\epsilon^{10}}\over{3\pi GM^2}}\int_0^{\infty} {\cal
F}(\nu)d\nu\ \ \ \ \text{with}\ \ \ \ {\cal
F}(\nu)\equiv{{4\nu^9+5\nu^7+\nu^5}\over{e^{4\pi\nu}-1}}
\end{equation}
for the characteristic radiation power of RN black holes in the
near-extremal regime (\ref{Eq1}).

From the analytical expression (\ref{Eq8}) for ${\cal F}(\nu)$ one
learns that the Hawking emission spectra of the near-extremal RN
black holes have a peak at the characteristic dimensionless
frequency \cite{Hodec}
\begin{equation}\label{Eq9}
\nu=\nu_{\text{peak}}\simeq 0.511\  .
\end{equation}
The corresponding energies of the emitted black-hole quanta are
characterized by the simple near-extremal relation [see Eqs.
(\ref{Eq1}), (\ref{Eq4}), and (\ref{Eq7})]
\begin{equation}\label{Eq10}
E=\hbar\omega=\hbar\nu_{\text{peak}}\sqrt{{{8\Delta}\over{M^3}}}\ .
\end{equation}
The quantum emission of the characteristic neutral Hawking field
mode (\ref{Eq10}) would produce a new spacetime configuration whose
mass and electric charge are given by [see Eq. (\ref{Eq1})]
\begin{equation}\label{Eq11}
M_{\text{new}}=M-E=Q+\Delta-E\ \ \ \ \text{and}\ \ \ \
Q_{\text{new}}=Q  .
\end{equation}

Intriguingly, one learns from Eq. (\ref{Eq11}) that the black-hole
condition $Q_{\text{new}}\leq M_{\text{new}}$ (and with it the
Penrose cosmic censorship conjecture \cite{Pen1,Pen2}) would be {\it
violated} due to the emission of the characteristic Hawking quanta
(\ref{Eq10}) from near-extremal RN black holes in the dimensionless
regime
\begin{equation}\label{Eq12}
\Delta <{{8(\hbar\nu_{\text{peak}})^2}\over{M^3}}\  .
\end{equation}

{\it Summary}. ---\ \ In the present compact essay we have analyzed
the Hawking emission spectra of charged Reissner-Nordstr\"om black
holes in the dimensionless near-extremal regime (\ref{Eq1}).
Interestingly, it has been shown that the semi-classical radiation
spectra of these near-extremal black holes can be studied
analytically in the large-mass regime (\ref{Eq2}).

We have explicitly proved that the characteristic Hawking emission
of quantum fields from black holes may turn an initially
near-extremal RN black hole with [see Eqs. (\ref{Eq1}), (\ref{Eq4}),
and (\ref{Eq12})] \cite{Notetem}
\begin{equation}\label{Eq13}
T_{\text{BH}}<{{2\nu_{\text{peak}}\hbar^2}\over{\pi M^3}}\
\end{equation}
into an horizonless naked singularity which is characterized by the
inequality $Q_{\text{new}}>M_{\text{new}}$ \cite{Hodec}. Our
analysis has therefore revealed the intriguing fact that the Hawking
evaporation process of black holes may violate the fundamental
Penrose cosmic censorship conjecture.

\newpage

\bigskip
\noindent
{\bf ACKNOWLEDGMENTS}
\bigskip

This research is supported by the Carmel Science Foundation. I thank
Don Page for interesting correspondence. I would also like to thank
Yael Oren, Arbel M. Ongo, Ayelet B. Lata, and Alona B. Tea for
stimulating discussions.



\begin{thebibliography}{99}
\bibitem{NoteQM} Here $M$ and $Q$ are respectively the gravitational mass and the
electric charge of the black hole. We shall assume $Q>0$ without
loss of generality.
\bibitem{Noteunit} We use natural units in which $G=c=k_{\text{B}}=1$.
\bibitem{Notepqm} Note that charged RN black-hole spacetimes are characterized
by the inequality $M\geq Q$ ($\Delta\geq0$), whereas over-charged
naked singularities, which are physically unacceptable according to
the Penrose cosmic censorship conjecture \cite{Pen1,Pen2}, are
characterized by the opposite relation $Q>M$.
\bibitem{Notel1} In particular, due to the presence of the centrifugal
barrier which surrounds the emitting black holes, the Hawking
radiation power of massless (electromagnetic and gravitational) fields
with larger values of the angular momentum parameter ($l>1$) is
suppressed by several factors of the large ratio $M/\Delta$ as
compared to the Hawking radiation power of unit angular momentum ($l=1$)
photons \cite{Pageev,Hodec}.
\bibitem{Noteerx} Here we have used the relations [see Eqs. (\ref{Eq1}) and (\ref{Eq7})]
$\epsilon=\sqrt{{{8\Delta}/{M}}}\cdot[1+O(\epsilon)]$ and
$\omega=\nu\sqrt{{{8\Delta}/{M^3}}}\cdot[1+O(\epsilon)]$ for the
near-extremal ($\epsilon\ll1$) black holes.
\end{thebibliography}
\end{document}